# Alternative Detectors for Spectrum Sensing by Exploiting Excess Bandwidth

Sirvan Gharib, Abolfazl Falahati and Vahid Ahmadi

Abstract- The problems regarding spectrum sensing are studied by exploiting a priori and a posteriori in information of the received noise variance. First, the traditional Average Likelihood Ratio (ALR) and the General Likelihood Ratio Test (GLRT) detectors are investigated under a Gamma distributed function as a channel noise, for the first time, under the availability of a priori statistical distribution about the noise variance. Then, two robust detectors are proposed using the exiting excess bandwidth to deliver a posteriori probability on the received noise variance uncertainty. The first proposed detector that is based on traditional ALR employs marginal distribution of the observation under available a priori and a posteriori of the received signal, while the second proposed detector employs the Maximum a posteriori (MAP) estimation of the inverse of the noise power under the same hypothesizes as the first detector. In addition, analytical expressions for the performance of the proposed detectors are obtained in terms of the false-alarm and detection probabilities. The simulation results exhibit the superiority of the proposed detectors over the traditional counterparts.

## I. Introduction

Wireless networks have a fixed spectrum assignment policy. The regulatory bodies such as Ofcom in the UK and the FCC in the USA have approved that portions of allocated spectrum bandwidths are underutilized. Therefore, the current fixed spectrum allocation policies may be inefficient [23]. To reduce underutilization of allocated fixed spectrum, cognitive radio networks (CRNs) were first proposed by J.Mitola in his PhD dissertation. It is an important requirement of CRNs that they sense portions of the spectrum that are not being used [1].

To enable secondary users (SUs) to utilize spectrum in an opportunistic manner, CRNs are sought to identify spectrum holes with sensing [2]-[3]. The energy detectors (ED) have become a popular sensing method due to low computational complexity and possible optimization in the presence of Additive White Gaussian Noise (AWGN) [7], [8]. However, the performance of the ED is susceptible to the errors from the channel added noise variance and deleterious effects of the channel, such as shadowing and multi path fading. Signal to Noise Ratio (SNR) must be above a certain threshold for achieving a specific detection probability under conditions of noise uncertainty [9]. One possible approach to alleviating detrimental channel impacts is to use cooperative spectrum sensing where some spatially distributed SUs are incorporated in sensing procedure [10]-[16]. Optimization of cooperative spectrum sensing over imperfect reporting channels with improved ED in each SU, is studied in [12]. Cooperative spectrum sensing for a CR mesh network is addressed in [13]. A linear cooperative sensing framework, based on the combination of observed energies by different SUs is proposed in [15]. In [16] and [17] , the authors propose a selective-relay based cooperative spectrum sensing scheme without dedicated reporting channel to control and reduce the interference from SUs to the primary user (PU). However, the performance of all the algorithms obtained in [7],[8],[12],[13],[15]-[17] substantially degrade in the presence of noise uncertainty.

The paper contributions are as follows: in this manuscript, our main contribution is to show even via uncertainty noise condition, the performance of energy detector can be improved. More specifically, we propose the square root raised cosine shaping filter to be employed by primary user. On the other hand, we recognized an accurate measuring in optimal mode. In this situation, this approach can be yield to robust detectors regarding noise variance uncertainty.

The paper organization: In section II, the basic assumptions and system model are described. In section III, the problem of the PU detection, using the ALR and GLR tests are taken into consideration to allow the readers to have a background for the proposed detectors. In section IV, the two new detectors that exploit a posteriori information are presented by Gamma distribution function. In section V, the performance of the proposed detectors are compared with the performance of the existing ALR and GLR detectors but with Gama Distribution. Simulation results are obtained to validate the results in Rayleigh and Nakagami fading channels to compare the performance of the investigated different detectors in section VI. Finally, section VII concludes the paper.

## II. BASIC ASSUMPTIONS AND SYSTEM MODEL

It is assumed that the PU utilizes raised cosine signal pulses with bandwidth B and roll off factor β, filtered and sampled with rate $T_s$ at the receiver, so that the SU receives N consecutive temporal samples. The hypotheses of the absence and presence of the PU signal are denoted by $H_0$ and $H_1$, respectively. Accordingly, the hypothesis testing problem for the detection of the PU signal can be considered as,

$$\begin{cases} H_0: z(n) = \eta(n) \\ H_1: z(n) = hs(n) + \eta(n) \end{cases} \quad n=1,\ldots,N \tag{1}$$

where $z(n)$, $s(n)$, $\eta(n)$ and $h$ denote the received signal, PU signal and noise samples at $n^{th}$ time instant, respectively and complex channel gain refers to Rayleigh and Nakagami fading. The PU signal is assumed to be distributed as a circular complex Gaussian random variable. As an example, this assumption is valid when the PU signal is considered to be modulated as an Orthogonal Frequency Division Multiplexing (OFDM) with sufficiently large number of sub-carriers [18]. In addition, it is also assumed that, the noise samples follow a circular complex Gaussian distribution. It is reasonable to assume that the observations of SUs are independently and identically distributed (i.i.d). In both hypothesis, the observations, $\mathbf{r} = [r(1),\ldots,r(N)]$ are taken as the squared envelopes of $z(n)$. Hence, the distributions of $\mathbf{r}$ under two hypothesis are,

$$f(r|\alpha,H_0) = \frac{1}{\alpha^N} exp\left(\frac{-N\bar{r}}{\alpha}\right) \tag{2}$$

$$f(r|\alpha,H_1) = \frac{1}{\alpha^N(1+\gamma)^N} exp\left(-\frac{N\bar{r}}{\alpha(1+\gamma)}\right) \tag{3}$$

where γ, α and $\bar{r}$ denote the SNR, noise power and the arithmetic mean of the observations, respectively. Alternatively, the observations can be taken as the magnitude-squared of the Fast Fourier Transform (FFT) of $z(n)$ referred by $w = [w(1),\ldots,w(N)]$. The normalized PSD of a PU signal is drawn as an example, with roll-off factor β=0.25 and bandwidth B=54KHZ and it is shown in figure 1.

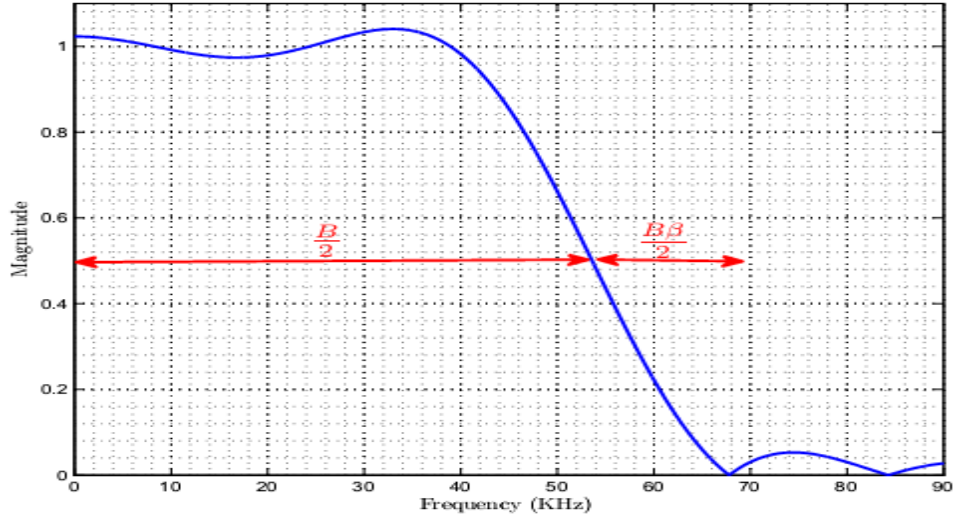

Fig. 1: The normalized PSD of a PU signal with roll-off factor β=0.25 and bandwidth B=54KHz

It can be noted that the frequency band $\pm\left[\frac{B}{2}, \frac{(1+\beta)B}{2}\right]$ consists most of noise information and the band $\left[-\frac{B}{2}, \frac{B}{2}\right]$ is composed of the PU signal plus noise information under $H_1$ [19], [20]. If there are L and P number of FFT samples within the bands $\left[-\frac{B}{2}, \frac{B}{2}\right]$, $\pm\left[\frac{B}{2}, \frac{(1+\beta)B}{2}\right]$ respectively, the observations vector, $w$, will be comprised of two vectors x=[x(1),...x(L)] and y=[y(1),...,y(P)], N=P+L, so that,

$$\begin{cases} H_0: \begin{cases} y(m) = v(m), & m = 1,\ldots,p \\ x(m) = v(m), & m = 1,\ldots,L \end{cases} \\ H_1: \begin{cases} y(m) = v(m), & m = 1,\ldots,P \\ x(m) = e(m) + v(m) & m = 1,\ldots,L \end{cases} \end{cases} \quad (4)$$

where $v(k)$ and $e(k)$ are magnitude-squared of the FFT of η(n) and s(n) at frequency $f_k$, respectively. Consequently, the distribution of w under each hypothesis is found as,

$$f(w|\alpha, H_0) = f(x, y|\alpha, H_0) \quad (5)$$
$$= \frac{1}{\alpha^N} \exp\left(\frac{-L\overline{x}}{\alpha}\right) \exp\left(\frac{-P\overline{y}}{\alpha}\right)$$

$$f(w|\alpha, H_1) = f(x, y|\alpha, H_1) = \frac{1}{\alpha^N(1+\gamma)^L} \exp\left(\frac{-L\overline{x}}{\alpha(1+\gamma)}\right) \exp\left(\frac{-P\overline{y}}{\alpha}\right) \quad (6)$$

Within this context, the well-known Gamma distribution for a random variable α is chosen to describe the noise power. It is assumed that, α is random variable with Gamma distribution,

$$f_\alpha(\alpha)=\frac{\theta}{k!}(\theta\alpha)^k exp(-\alpha\theta)u(\alpha) \qquad (7)$$

where $u(\alpha)$ is the Heaviside step function and the parameters $k$ and $\theta$ should be chosen so that, the given distribution in (7) is in a close agreement with the actual distribution of the noise power. If the power of noise ($\alpha$) is assumed to be known, the optimum detector in Neyman-Pearson sense is found as [24],

$$L(x) = \frac{f(x|\alpha, H_1)}{f(x|\alpha, H_0)} \gtrless_{H_0}^{H_1} \tau \qquad (8)$$

Since (8) is a strictly monotonous function (i.e, monotonically increasing like logarithm function) of $\bar{x}$, it is possible to simply compare $\bar{x}$ with a threshold for decision making to yield the detector as,

$$T_{opt} = \bar{x} \gtrless_{H_0}^{H_1} \eta \qquad (9)$$

where $\eta$ is the decision threshold.

## III. TRADITIONAL DETECTORS

In this section, the temporal samples of the received signal **r**, is employed as observations for spectrum sensing. In the following subsections, the corresponding detectors are acquired based on the ALR and GLR detectors using a Gamma distributed function as a channel noise, with the condition of availability of a priori statistical distribution about the noise variances.

### A. Gamma Distributed ALR Detector

When the prior information about an unknown parameter is available, the traditional approach is to use the ALR detector. This test requires the marginal distribution of the observation under both hypothesis, which can be computed as follows,

$$f(r|H_0) = \int_0^\infty f(r|\alpha, H_0)f_\alpha(\alpha)d\alpha \qquad (10)$$
$$= \frac{\theta^{k+1}}{k!} \frac{(N+k)!}{(\theta + N\bar{r})^{N+k+1}}$$

$$f(r|H_1) = \int_0^\infty f(r|\alpha, H_1) f_\alpha(\alpha) d\alpha = \frac{\theta^{k+1}}{k!(1+\gamma)^N} \frac{(N+k)!}{(\theta + \frac{N\bar{r}}{(1+\gamma)})^{N+k+1}} \qquad (11)$$

Constituting the likelihood ratio leads to following decision statistics,

$$L(r) = \frac{f(r|H_1)}{f(r|H_0)} = \frac{1}{(1+\gamma)^N} \left( \frac{1 + \frac{N\bar{r}}{\theta}}{1 + \frac{N\bar{r}}{\theta(1+\gamma)}} \right) \gtrless_{H_0}^{H_1} \tau \qquad (12)$$

Considering L(r) as a monotones function (i.e, absolutely linear ascending) of $\bar{r}$, the decision statistic can be simplified as,

$$T_{ALRD1} = \frac{N\bar{r}}{\theta} \gtrless_{H_0}^{H_1} \eta \qquad (13)$$

The function $T_{ALRD1}$ is the Average Likelihood Ratio Decision.

**B. Gamma Distributed GLR Detector**

An alternative approach is to use GLR that employs the Maximum A Posteriori (MAP) estimate of the inverse of noise power $\alpha$. The MAP estimate of α under hypothesis $H_i$ is obtained as,

$$\hat{\alpha}_i = argmax\, f_\alpha(\alpha) f(r|\alpha, H_i) \qquad (14)$$

Hence, the absence and presence of the signal can be computed as,

$$\hat{\alpha}_0 = argmax\, \frac{\theta^{k+1}}{k!} \frac{1}{\alpha^{N+k}} exp\left(-\frac{(\theta + N\bar{r})}{\alpha}\right) \qquad (15)$$

$$\hat{\alpha}_1 = argmax\, \frac{\theta^{k+1}}{k!} \frac{1}{\alpha^{N+k}(1+\gamma)^N} exp\left(-(\theta + \frac{N\bar{r}}{1+\gamma})\frac{1}{\alpha}\right) \qquad (16)$$

which leads to,

$$\alpha_0 = \frac{\theta + N\bar{r}}{N + k} \tag{17}$$

$$\alpha_1 = \frac{\theta + \frac{N\bar{r}}{1+\gamma}}{N + k} \tag{18}$$

Replacing the estimates (17) and (18) with (10) and (11), respectively and forming likelihood ratio, it is found that,

$$L_{GLRD1}(r) = \frac{f(r|\hat{\alpha}_1, H_1)}{f(r|\hat{\alpha}_0, H_0)} \tag{19}$$

$$= \left(\frac{1 + \frac{N\bar{r}}{\theta}}{1 + \gamma + \frac{N\bar{r}}{\theta}}\right)^N exp(\gamma(N+k) \frac{\frac{N\bar{r}}{\theta}}{(1+\frac{N\bar{r}}{\theta})(1+\gamma+\frac{N\bar{r}}{\theta})}) \gtrless_{H_0}^{H_1} \eta$$

The function $L_{GLRD1}(r)$ is the Generalized Likelihood Ratio Decision, and a function of $\frac{N\bar{r}}{\theta}$ that has one only extreme at $\mu \triangleq \frac{N(2+\gamma)+\sqrt{(2+\gamma)^2 N^2 + 4k(1+\gamma)(2N+k)}}{2k}$ and is a maxima. Thus, $L_{GLRD1}$ is a monotonically increasing function for $\frac{N\bar{r}}{\theta} < \mu$ and is a monotonically decreasing function for $\frac{N\bar{r}}{\theta} > \mu$. It is possible to simply compare $\eta$ with a threshold for decision making to yield the detector as,

$$T_{GLRD1}(r) = \begin{cases} \frac{N\bar{r}}{\theta} \gtrless_{H_0}^{H_1} \eta_1 \ if \ \frac{N\bar{r}}{\theta} < \mu \\ \frac{N\bar{r}}{\theta} \gtrless_{H_0}^{H_1} \eta_2 \ if \ \frac{N\bar{r}}{\theta} > \mu \end{cases} \tag{20}$$

It can be shown that $\eta_1 < \mu < \eta_2$, which leads to the following decision rule,

$$\begin{cases} H_0: if \ \frac{N\bar{r}}{\theta} < \eta_1 \ or \ \frac{N\bar{r}}{\theta} > \eta_2 \\ H_1: if \ \eta_1 < \frac{N\bar{r}}{\theta} < \eta_2 \end{cases} \tag{21}$$

## IV. PROPSED DETECTORS

In this section, samples of the received signal in frequency domain, $w$ are exploited as observations to sense spectrum. Under such conditions, the part of $w$ that is compromised of the noise information, y can be used to achieve a posteriori information about the noise power. For this purpose, using Bayes' s law, it can be written as follows,

$$f_\alpha(\alpha|y) = \frac{f(y|\alpha)f_\alpha(\alpha)}{\int_0^\infty f(y|\alpha)f_\alpha(\alpha)d\alpha} \qquad (22)$$

$$= \frac{\frac{1}{\Gamma(k)}\theta^k \alpha^{-(P+k-1)} exp\left(-\frac{(\theta+P\bar{y})}{\alpha}\right)}{\frac{1}{\Gamma(k)}\theta^k \int_0^\infty \alpha^{-(P+k-1)} exp\left(-\frac{(\theta+P\bar{y})}{\alpha}\right)d\alpha}$$

$$= \frac{(\theta+P\bar{y})^{P+k+1} \alpha^{-(P+k)}}{(P+k)!} exp\left(-\frac{(\theta+P\bar{y})}{\alpha}\right)$$

where $\bar{y}$ denotes the arithmetic mean of y.

### A. Proposed ALR Detector

The data-aided conditional probability distribution of x under each hypothesis can be computed as,

$$f(x|y, H_0) = \int_0^\infty f(x|\alpha, H_0)f_\alpha(\alpha|y)d\alpha \qquad (23)$$

$$= \frac{(\theta+P\bar{y})^{P+k+1}}{(P+k)!} \frac{(P+L+k)!}{(\theta+P\bar{y}+L\bar{x})^{P+L+k+1}}$$

$$f(x|y, H_1) = \int_0^\infty f(x|\alpha, H_1)f_\alpha(\alpha|y)d\alpha \qquad (24)$$

$$= \frac{(\theta+P\bar{y})^{P+k+1}}{(1+\gamma)^L(P+k)!} \frac{(P+L+k)!}{\left(\theta+P\bar{y}+\frac{L\bar{x}}{1+\gamma}\right)^{P+L+k+1}}$$

The relations (23) and (24) are obtained using the independence of the random vector x and y with a condition that the parameter $\alpha$ is known. After some manipulation, the likelihood ratio can be determined as follows,

$$L(x|y) = \frac{1}{(1+\gamma)^L} \left( \frac{1 + \frac{L\bar{x}}{\theta + P\bar{y}}}{1 + \frac{L\bar{x}}{(\theta + P\bar{y})(1+\gamma)}} \right)^{P+L+k+1} \gtreqless_{H_0}^{H_1} \eta \tag{25}$$

Since the above function is a monotonous function of $\frac{L\bar{x}}{\theta + P\bar{y}}$, the decision statistic can be simplified as,

$$T_{ALRD2}(x, y) = \frac{L\bar{x}}{\theta + P\bar{y}} \gtreqless_{H_0}^{H_1} \eta \tag{26}$$

### B. Proposed GLR Detector

As mentioned in III-B, an alleviative method is to use GLR. According to (23) and (24), it can be found that,

$$\hat{\alpha}_0 = argmax \frac{(\theta + P\bar{y})^{k+P+1}}{(k+P)!} \alpha^{L+k+P} exp(-(\theta + P\bar{y} + L\bar{x})\alpha) \tag{27}$$

$$\hat{\alpha}_1 = argmax \frac{(\theta + P\bar{y})^{k+P+1}}{(k+P)!} \frac{\alpha^{L+k+P}}{(1+\gamma)^L} exp\left(-\left(\theta + P\bar{y} + \frac{L\bar{x}}{1+\gamma}\right)\alpha\right) \tag{28}$$

Hence, the following estimates can be performed as,

$$\hat{\alpha}_0 = \frac{\theta + P\bar{y} + L\bar{x}}{L + k + P} \tag{29}$$

$$\hat{\alpha}_1 = \frac{\theta + P\bar{y} + \frac{L\bar{x}}{1+\gamma}}{L + k + P} \tag{30}$$

Statistically, the above estimates of the likelihood ratio will be

$$L_{GLRD2}(x) = \frac{f(x|\hat{\alpha}_1, H_1)}{f(x|\hat{\alpha}_0, H_0)} = \left( \frac{1 + \frac{L\bar{x}}{\theta + P\bar{y}}}{1 + \gamma + \frac{L\bar{x}}{\theta + P\bar{y}}} \right)^L exp(\gamma(L + k + P) \frac{\frac{L\bar{x}}{\theta + P\bar{y}}}{(1 + \frac{L\bar{x}}{\theta + P\bar{y}})(1 + \gamma + \frac{L\bar{x}}{\theta + P\bar{y}})} \gtreqless_{H_0}^{H_1} \eta \tag{31}$$

$L_{GLRD2}(x)$ is a function of $\frac{L\bar{x}}{\theta+P\bar{y}}$ which has only one extreme in $\rho \overset{\Delta}{=} \frac{L(2+\gamma)+\sqrt{(2+\gamma)^2 L^2+4(k+P)(1+\gamma)(2L+k+P)}}{2(k+P)}$. Similar to III-B, the decision rule is,

$$\begin{cases} H_0: if \; \frac{L\bar{x}}{\theta+P\bar{y}} < \eta_1 \; or \; \frac{L\bar{x}}{\theta+P\bar{y}} > \eta_2 \\ H_1: if \; \eta_1 < \frac{L\bar{x}}{\theta+P\bar{y}} < \eta_2 \end{cases} \qquad (32)$$

## V. PERFORMANCE EVALUATION

In this section, the performance of the proposed detectors in the presence of noise and fading are evaluated in terms of the detection and false-alarm probabilities, i.e. $P_d$ and $P_{fa}$ in order to enable examining their robustness to noise uncertainty.

### A. Performance of Optimal Detector

Amongst the aforementioned detectors, the optimum detector has the simplest structure, but requires knowledge of the noise power. The decision statistic of the optimum detector is composed of the summation of N exponentially distributed random variables. This leads to a Gamma distribution with N degrees of freedom. Consequently, $P_{fa}$ and $P_d$ are acquired as follows [22],

$$P_{fa} = \frac{\Gamma\left(N, \frac{\eta}{\alpha}\right)}{\Gamma(N)} \qquad (33)$$

$$P_d = \frac{\Gamma\left(N, \frac{\eta}{\alpha(1+\gamma)}\right)}{\Gamma(N)} \qquad (34)$$

where $\Gamma(s,x) = \int_x^\infty t^{s-1} e^{-t} dt$ is upper incomplete Gamma function.

### B. Performance of Traditional Detectors

According to (13), the decision statistics of the traditional ALR detector is a scaled version of the optimal detector. The scaled version of a Gamma random variable, $\frac{x}{\alpha}$ with the shape parameter k and inverse scale parameter $\theta$ will also follow a Gamma distribution with the shape parameter k and inverse scale parameter $\frac{\theta}{\alpha}$.

Accordingly, $P_{fa}$ and $P_d$ of the traditional ALR detector are computed as,

$$P_{fa}(\alpha) = \frac{\Gamma\left(N, \frac{\theta\eta}{N\alpha}\right)}{\Gamma(N)} \tag{35}$$

$$P_d(\alpha) = \frac{\Gamma\left(N, \frac{\theta\eta}{N\alpha(1+\gamma)}\right)}{\Gamma(N)} \tag{36}$$

where α is noise power and has gamma distribution and h is channel gain.

To evaluate the performance of traditional GLR detector, the following terms must be calculated,

$$P_{fa}(\alpha) = P(H_1|H_0) = P\left[\eta_1 < \frac{N\bar{r}}{\theta} < \eta_2 \Big| H_0\right] \tag{37}$$

$$P_d(\alpha) = P(H_1|H_1) = P\left[\eta_1 < \frac{N\bar{r}}{\theta} < \eta_2 \Big| H_1\right] \tag{38}$$

It can be well demonstrated that $\frac{N\bar{r}}{\theta}$ has a statistical average which is always less than μ. Considering Markov's inequality, it can be shown that $P\left[\frac{N\bar{r}}{\theta} > \mu\right]$ is negligible and tends to zero, particularly in a low SNR regime. This property is of interest in CRNs. Therefore, we usually have $\frac{N\bar{r}}{\theta}$. Hence we can write,

$$P_{fa}(\alpha) = P\left[\eta_1 < \frac{N\bar{r}}{\theta} < \mu \Big| H_0\right] = 1 - P\left(\frac{N\bar{r}}{\theta} < \eta_1 \Big| H_0\right) \tag{39}$$
$$= P\left(\frac{N\bar{r}}{\theta} > \eta_1\right) = \frac{\Gamma\left(N, \frac{\theta\eta_1}{N\alpha}\right)}{\Gamma(N)}$$

$$P_d(\alpha) = P\left[\eta_1 < \frac{N\bar{r}}{\theta} < \mu \Big| H_1\right] = 1 - P\left(\frac{N\bar{r}}{\theta} < \eta_1 \Big| H_1\right) \tag{40}$$
$$= P\left(\frac{N\bar{r}}{\theta} > \eta_1\right) = \frac{\Gamma\left(N, \frac{\theta\eta_1}{N\alpha(1+\gamma)}\right)}{\Gamma(N)}$$

Thus, averaging probabilities are needed to be evaluated by averaging already derived probabilities over the α varying statistic. Hence,

$$\overline{P_d} = \int_\alpha P_d(\alpha) f_\alpha(\alpha) d\alpha \tag{41}$$

$$\overline{P_{fa}} = \int_\alpha P_{fa}(\alpha) f_\alpha(\alpha) d\alpha \tag{42}$$

where $f_\alpha(\alpha)$ represents the pdf of noise power.

## C. Performance of Proposed Detectors

To compute $P_{fa}$ and $P_d$, the decision statistic in (26) algorithm can be rewritten as,

$$\Phi = L\bar{x} - P\eta\bar{y} \gtrless_{H_0}^{H_1} \theta \tag{43}$$

where $\bar{x}$ and $\bar{y}$ denote the arithmetic mean of the observation in the frequency domain. If we have N samples of noise with normal distribution, we will have N samples of noise with normal distribution after taking the Fourier transform.

Since $\phi$ is the sum of several Gamma variables that are complex to express, the central limit theory (CLT) can be used as long as the number of observations is large enough. With the aid of Appendix A, $P_{fa}$ and the average of $P_d$ for fading channels can be obtained as,

$$P_{fa}(\alpha) = Q\left(\frac{\theta\eta - \alpha N(L - P\eta)}{\alpha N \sqrt{L + P\eta^2}}\right) \tag{44}$$

$$P_d(\alpha, h_i, h_r, s_i, s_r) \tag{45}$$
$$= Q\left(\frac{\theta\eta - (L((h_r s_r - h_i s_i)^2 + (h_i s_r + h_r s_i)^2 + N\alpha) + NP\eta\alpha)}{\sqrt{2NL\alpha((h_r s_r - h_i s_i)^2 + (h_i s_r + h_r s_i)^2 + N\alpha) + N^2 P\eta^2 \alpha^2}}\right)$$

where $P_{fa}$ is a function of $\alpha$ and $P_d$ is a function of $\alpha$, $h_r, h_i, s_r$ and $s_i$. Thus, averaging probabilities are needed to be evaluated by averaging already derived probabilities over the varying statistics. Hence,

$$\overline{P_{fa}} = \int_\alpha P_{fa}(\alpha) d\alpha \tag{46}$$

$$\overline{P_d} = \int_{\alpha, h_i, h_r, s_i, s_r} P_d(\alpha, h_i, h_r, s_i, s_r) d\alpha dh_i dh_r ds_i ds_r \qquad (47)$$

where $\alpha, h_r, h_i, s_r$ and $s_i$ are the inverse of the noise power, the real part of the channel gain, the imaginary part of the channel gain, the real part of the primary user signal and the imaginary part of the primary user signal, respectively.

For the GLRT detector, similar calculations as section V part B are followed as,

$$P_{fa}(\alpha) = P(H_1|H_0) = P\left[\eta_1 < \frac{L\bar{x}}{\theta + P\bar{y}} < \eta_2 \Big| H_0\right] \qquad (48)$$

$$P_d(\alpha, h_i, h_r, s_i, s_r) = P(H_1|H_1) \qquad (49)$$
$$= P\left[\eta_1 < \frac{L\bar{x}}{\theta + P\bar{y}} < \eta_2 \Big| H_1\right]$$

It can be well demonstrated that $\frac{L\bar{x}}{\theta + P\bar{y}}$ has a statistical average which is always less than ρ, using *Markov's* inequality,

$$P_{fa}(\alpha) = P\left[\eta_1 < \frac{L\bar{x}}{\theta + P\bar{y}} < \rho \Big| H_0\right] = 1 - P\left(\frac{L\bar{x}}{\theta + P\bar{y}} < \eta_1 \Big| H_0\right) \qquad (50)$$
$$= P\left(\frac{L\bar{x}}{\theta + P\bar{y}} > \eta_1 \Big| H_0\right)$$

$$P_d(\alpha, h_i, h_r, s_i, s_r) = P\left[\eta_1 < \frac{L\bar{x}}{\theta + P\bar{y}} < \rho \Big| H_1\right] = 1 - P\left(\frac{L\bar{x}}{\theta + P\bar{y}} < \eta_1 \Big| H_1\right) = \qquad (51)$$
$$P\left(\frac{L\bar{x}}{\theta + P\bar{y}} > \eta_1 \Big| H_1\right)$$

Relationships (50) and (51) lead to relations (44) and (45). It can be realized that, the performance of ALR and GLR detectors are about the same, hence simulation studies are performed for ALR detector only.

To compare the traditional and proposed methods, the statistical properties of the distributions should be examined. Below are the statistical properties of both methods for when user is present (appendix A). Note that in both methods, assuming that the number of samples is sufficient, the CLT approximation is used.

$$\Lambda_{Traditional} = \mathcal{N}(N\alpha(1 + \gamma), N\alpha^2(1 + \gamma)^2) \qquad (52)$$

$$\Lambda_{Proposed} = \mathcal{N}\left(2NL\alpha(2\gamma + 1 - P\eta), 8LN^2\alpha^2\left(2\gamma + \frac{1}{2} + \frac{P\eta^2}{2}\right)\right) \quad (53)$$

## VI. Simulation Results

In this section, we intend to analyze the results obtained in the previous sections by providing analytical relations simulations as well as numerical simulations. To perform numerical simulations, the Monte Carlo algorithm is employed in which, for each obtained point two sets of results are derived, one 20*10000 iterations and one for 40*10000 iterations, the values of 20 and 40 represent the number of observations each with 10000 samples. we can also observe that as N increases, the performance of the detectors improves. This increase improves the performance of the proposed detectors more than the traditional detectors.

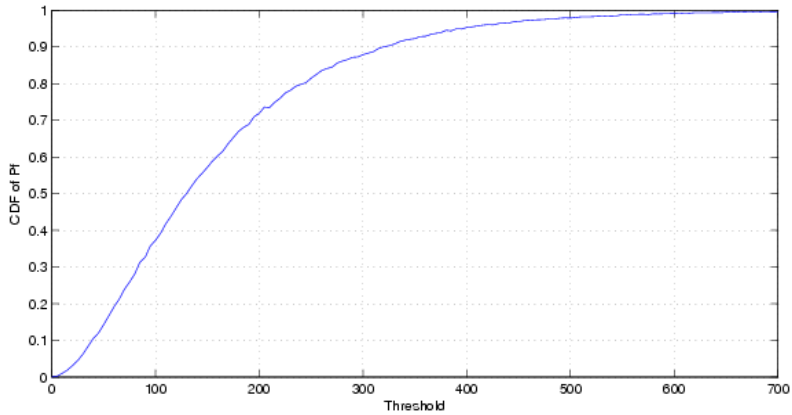

Fig. 2: CDF of false alarm distribution function for traditional method at N=20

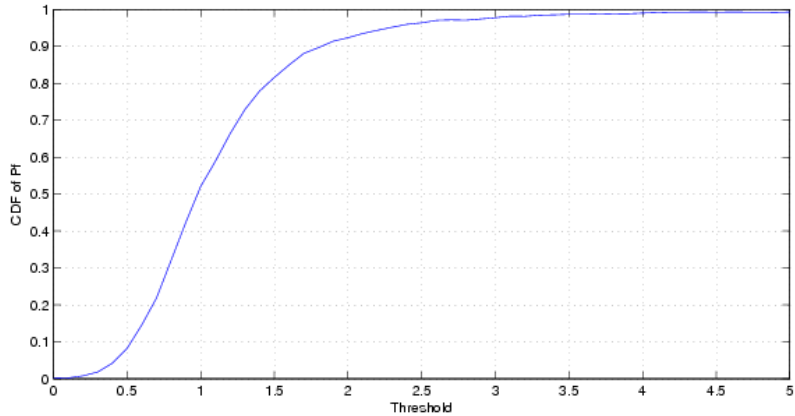

Fig. 3: CDF of false alarm distribution function for proposed method at N=20

In comparing the conventional with proposed method, we must pay attention to decision metrics. In conventional method, decision metric is the energy of observations in the time domain, and in the proposed method, decision metric is the energy of first part minus second part in the frequency domain. Of course, it should be noted that the threshold coefficient multiplied by the energy of second part.

Figures 2 and 3 show the cumulative distribution function of the false alarm probability for the traditional and proposed modes. In the proposed case, it has a lower threshold level than in the traditional case, because in the proposed case, the decision metric is a subtraction of energy from one part of the other. Due to the energy back factor of the second part, with a slight increase in the threshold, the cumulative distribution function moves to the value of one more quickly.

Figure 4 shows the performance of optimal, traditional and proposed detectors in an AWGN channel with SNR at 0dB. In the optimal detector, the variance of the noise is certain but with uncertainty condition, its performance is greatly reduced. However, in the proposed detector with noise uncertainty, the proposed detector performs better than the traditional detector. The region of complementary (ROC) curve is plotted for 20 and 40 observations.

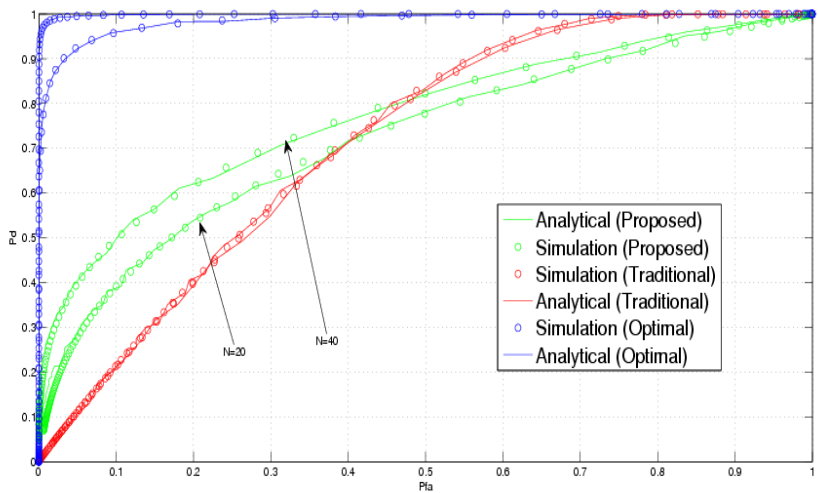

Fig. 4: The ROC curve of signal model over the AWGN channel with average SNR at 0dB and N=20,40

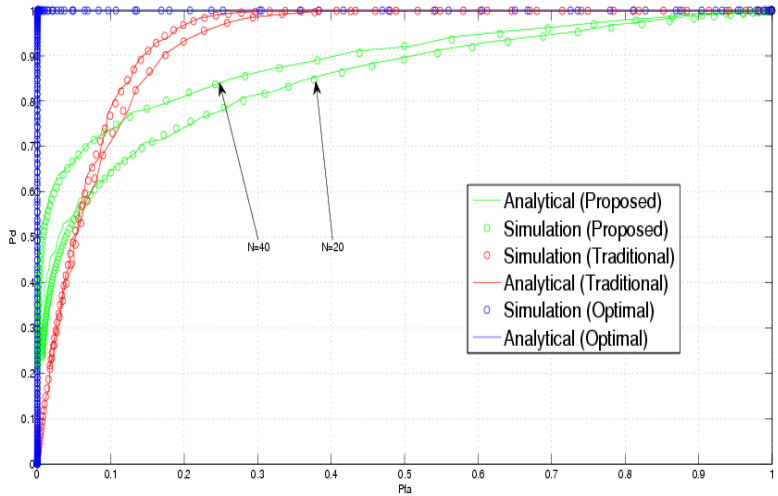

Fig. 5: The ROC curve of signal model over the AWGN channel with average SNR at 5dB and N=20,40

Figure 5 shows the performance of optimal, traditional and proposed detectors in an AWGN channel with SNR at 5dB. As can be seen, the performance improves with increasing SNR. It is worthy to mentioned that, at small values of $P_{fa}$, the proposed detectors still perform better than the traditional detectors.

Figure 6 shows the performance of optimal, traditional and proposed detectors in the Nakagami and Rayleigh channels SNR at 0dB. In this case, the detector performance improves with increasing N, m, and the ROC curve moves further to the top left corner.

Figure 7 shows the performance of optimal, traditional and proposed detectors in the Nakagami and Rayleigh channels with SNR at 5dB. In this particular case, at small amounts of $P_{fa}$, the proposed detectors perform better than the traditional detectors.

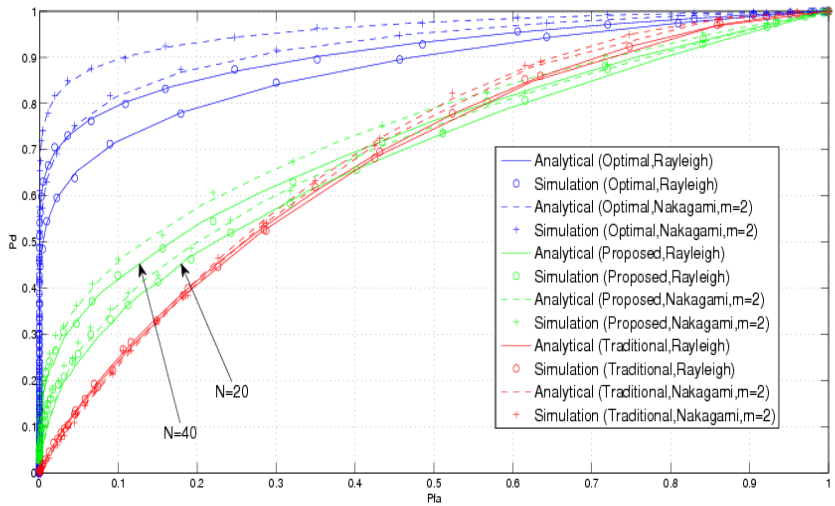

Fig. 6: The ROC curve of signal model over the Rayleigh and Nakagami channel with average SNR at 0dB and N=20,40

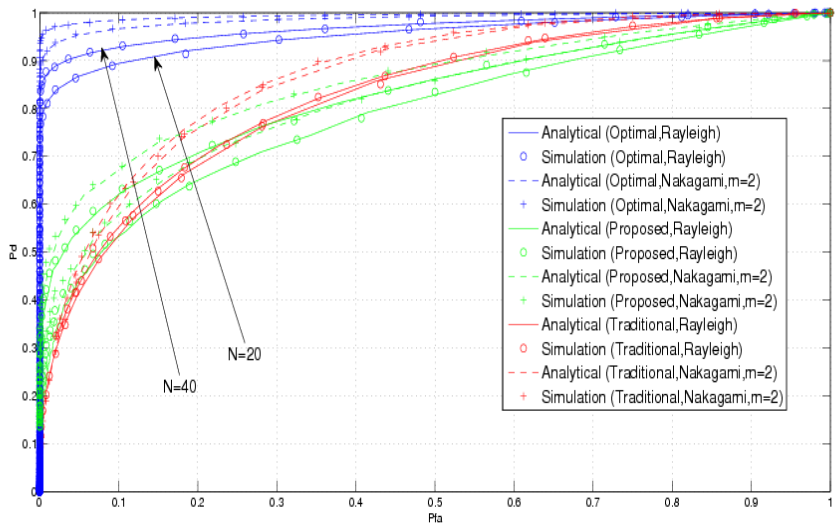

Fig. 7: The ROC curve of signal model over the Rayleigh and Nakagami channel with average SNR at 5dB and N=20,40

As the simulations show, the proposed method performs better in low probability false alarm. If we pay attention to (43), the decision metric is in the form of energy subtraction of the first and second parts by applying the threshold coefficient in the energy of the second part. According to the decision metric, the lower the threshold level, the energy of the second part is multiplied by a smaller number, and when the threshold is zero, the subtraction of the energy of the second part become ineffective, and if the threshold is negative, it will have a worse effect. In detection problem, the lower the threshold, the higher the area under the distribution probability function.

Furthermore, increasing the SNR improves the performance of the detectors, but this increase is greater in the traditional detector than in the proposed detector. According to (52) and (53), the distribution variance in the traditional detector corresponds to the power of 2 of the SNR, while in the proposed detector, the distribution variance corresponds to the SNR. As the number of samples increases, the performance of the detectors improves, but the proposed detector improves more than the traditional detector. This is because the distribution variance in the proposed detector includes a higher degree of sample size.

## VII. Conclusions

The new detectors for spectrum sensing in cognitive radio networks are proposed based on the ALR and GLR tests, under Gamma distributed noise power scenarios to exploit the availability of the excess bandwidth. The analytical expressions are obtained for the false-alarm and detection probabilities of the proposed detectors and are validated by simulation results. In addition, the provided simulation results revealed the superiority of the proposed detectors over their counterparts in practical scenarios. Furthermore, the speed of operations to perform the two proposed detectors (26) were found to be almost the same as traditional ones (13) such as ALR and GLR detectors.

## Appendix A

As stated earlier, we assume that the primary user uses the pulse shape of the raised cosine signal. Therefore, the frequency spectrum of the user is divided into two parts. In the first part, when the primary user is present $(H_1)$, the first L sample contains the user plus the noise and the second sample P contains only the noise. In the second part, when the primary user is absent $(H_0)$, all samples are noise related. As previously mentioned, the decision metric is expressed as $\phi = L\bar{x} - P\eta\bar{y} = \sum_1^L x - \eta \sum_1^P y$. Using the central limit theorem we can write,

$$H_0: x \sim y \sim N(N\alpha, N\alpha^2), P_{fa}(\alpha) = Q\left(\frac{\theta\eta - \alpha N(L - P\eta)}{\alpha N\sqrt{L + P\eta^2}}\right)$$

$H_1: y \sim N(N\alpha, N\alpha^2)$,
$x = |hs + n|^2 = |(h_R + jh_I)(s_R + js_I) + (n_R + jn_I)|^2 = (h_R s_R - h_I s_I + n_R)^2 + (h_I s_R + h_R s_I + n_I)^2$,

$$x = N(h_I^2 s_I^2 + h_I^2 s_R^2 + h_R^2 s_I^2 + h_R^2 s_R^2 + N\alpha)$$

$P_d(\alpha, h_i, h_r, s_i, s_r)$
$$= Q\left(\frac{\theta\eta - (L((h_r s_r - h_i s_i)^2 + (h_i s_r + h_r s_i)^2 + N\alpha) + NP\eta\alpha)}{\sqrt{2NL\alpha((h_r s_r - h_i s_i)^2 + (h_i s_r + h_r s_i)^2 + N\alpha) + N^2 P\eta^2\alpha^2}}\right)$$